# Be More Real: Travel Diary Generation Using LLM Agents and Individual Profiles

Xuchuan Li
Department of Urban Informatics
Shenzhen University
Shenzhen, China

Fei Huang
Department of Urban Informatics
Shenzhen University
Shenzhen, China

Jianrong Lv
Department of Urban Informatics
Shenzhen University
Shenzhen, China

Zhixiong Xiao
Department of Urban Informatics
Shenzhen University
Shenzhen, China

Guolong Li
Department of Urban Informatics
Shenzhen University
Shenzhen, China

Yang Yue [†]
Deartment of Urban Informatics
Shenzhen University
Shenzhen, China

**ABSTRACT**

Human mobility is inextricably linked to social issues such as traffic congestion, energy consumption, and public health; however, privacy concerns restrict access to mobility data. Recently, research have utilized Large Language Models (LLMs) for human mobility generation, in which the challenge is how LLMs can understand individuals' mobility behavioral differences to generate realistic trajectories conforming to real world contexts. This study handles this problem by presenting an LLM agent-based framework (MobAgent) composing two phases: understanding-based mobility pattern extraction and reasoning-based trajectory generation, which enables generate more real travel diaries at urban scale, considering different individual profiles. MobAgent extracts reasons behind specific mobility trendiness and attribute influences to provide reliable patterns; infers the relationships between contextual factors and underlying motivations of mobility; and based on the patterns and the recursive reasoning process, MobAgent finally generates more authentic and personalized mobilities that reflect both individual differences and real-world constraints. We validate our framework with 0.2 million travel survey data, demonstrating its effectiveness in producing personalized and accurate travel diaries. This study highlights the capacity of LLMs to provide detailed and sophisticated understanding of human mobility through the real-world mobility data.

## 1. INTRODUCTION

Human mobility behaviors are intricately linked to personal preferences, social interactions, and environmental factors, the parsing of mobility data while extracting accurate mobility patterns is essential for urban planning and traffic management[1, 29]. Although human mobility is regular and predictable, it is also inherently complex under the space-time and social constrains[8, 24, 25]. For human mobility generation, it is essentially attempting to understand and model how various factors such as individual's preferences, social attributes, and environmental conditions influence the mobility behaviors[5].

Deep learning methods, such as VAE[10, 16], GAN[4, 12], and Diffusion models[3, 33] have made considerable progress for generating individual trajectories. However, they are limited in understanding the intentional and contextual aspects of human mobility, especially when faced with atypical or nuanced scenarios, which hinders their effectiveness in generating realistic and contextually rich travel diaries. Large language models (LLM) have shown emergent abilities for semantic interpreting and generalized purpose reasoning, which allows it to simulate human behaviors with a narrowly defined context[2, 18] by reformulating the trajectory generation problems into a reasoning problem, and thus, brings opportunities for generating more realistic individual trajectories. LLMs are expected to summarize and reason personal mobility patterns and motivations, and then generating customized behavior based on these templates [22, 26]. However, most of the existing studies fail to take into account personal profiling, such as gender, age, and socioeconomic status, resulting in unrealistic trajectories lack individual characteristics. As illustrated in Figure 1, individuals with the same occupation but varying ages exhibit distinct patterns of mobility over time. However, when all "technical employees" are considered as a homogeneous group, the generated trajectories may only reflect the overall mobility tendencies associated with their occupational identity, failing to capture their personal differences. Thus, the new paradigm for mobility generation using LLMs faces the challenge of accurately understanding the complex relationship between individual attributes and mobility tendencies, and generating more realistic trajectories that are not only statistically plausible but also consistent with real-world contexts.

This study attempts to address the following research questions: RQ 1: How do LLMs align individual attributes with motivations and behaviors of human daily mobility. RQ 2: Can LLMs generate more real and personalized travel diaries in urban contexts. RQ 3: What is the impact of individual attributes on the performance of LLMs.

To answer the questions, we propose a fine-grained individual travel diary generation framework, *MobAgent*, based on LLM agents, which consists of two parts: (1) understanding-based mobility pattern extraction, and (2) reason-based mobility generation. In the first part, LLM is intend to learn the fundamental principle behind mobility. LLM reveals the relationships between personal attributes and mobility behaviors through a comparison and self-evaluation mechanism; such that extracting thoughtful mobility patterns after understanding how and why mobility behavior occurs. In the Second part, we introduce a recursive rethink strategy to reason individual mobility motivation along with the retrieval information in Part 1; that facilitate the generation of consecutive daily trajectory with

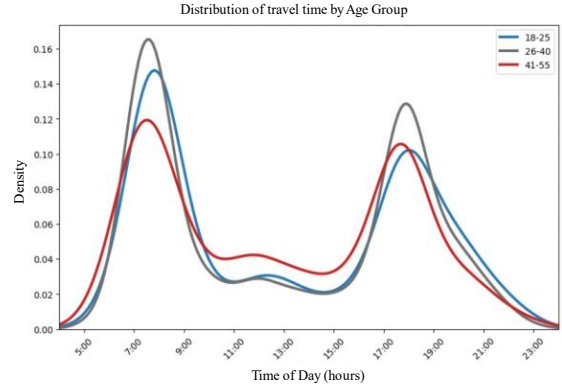

**Figure1**. Distribution of travel time among technical employees under different age groups.

representative individual attributes. Then, the generated trajectories are mapped into a real urban context considering the shortest path based on road network.

We evaluate the proposed framework using GPT-4.0 and DeepSeek APIs on a 0.2 million real-world travel survey data with individual profiles. The results demonstrate that the proposed framework, MobAgent, excels in generating personalized and context-aware travel diaries over the space-time and semantic scale. We summarize our contribution as follows: (1) This study reveals the dependency between individual attributes and mobility behaviors and proves the feasibility of LLM agents for generating more realistic individual trajectories. (2) We introduce an understanding-based patterns extraction mechanism, which supports the LLM to figure out the fundamental principle behind personalized mobility behaviors. (3) We propose a reason-based rethink strategy that generates interpretable motivation and mobility with considering road network structures in physical environment, which is accurately matched with individual's attributes.

## 2. RELATE WORKS

### 2.1 Mobility Data Generation

Early mobility modeling has relied on statistical methods, such as Markov model[17, 30], which captured sequential dependencies in mobility patterns but fell short in representing the multi-faceted dimensions of human behavior. The emergence of generative deep learning models [9, 6, 10] marked a turning point in the field. Such as VAEs introduced the concept of a latent space that allows for the generation of diverse trajectories, capturing a wide range of human movements[10]. However, they often struggled with interpretability and sensitivity to contextual factors. GANs have been particularly notable for their ability to generate realistic and diverse data samples. Feifei Li's team [7] released a model called "Social GAN" to synthesize human movement trajectories with social attributes for robotic platforms, which led to research on human trajectory prediction in related fields[12, 20, 21]. However, they also face challenges such as mode collapse and the need for large amounts of training data. DDPMs, more recently, have shown promise in producing high-quality synthetic data by modeling the data distribution through iterative denoising processes. The step-by-step modeling process make its superiority of capturing spatial-temporal features derived from real trajectories[31–33]. In addition, the combination of statistic model and deep learning model has been purposed as a more robust framework for capturing the complex, multi-dimensional nature of human mobility[4, 23, 27]. However, the randomness, sparsity, and irregular patterns in mobility are challenging.

## 2.2 LLM Agents and Human Behavior

LLMs possessing extensive knowledge and reasoning capabilities have recently been leveraged to replicate real-world dynamics to generate believable behavior[14, 18]; such as utilizing agents based on LLMs to simulate human decision-making in fields of epidemiology[28], sociology[19], and economics[14]. For instance, utilizing LLMs agents to simulate individual emotions, attitudes, and interactive behaviors, in a simulation social network environment[15]. As well as the treating illness scenario, all patients, nurses, and doctors are autonomous agents to simulate disease onset and progression, promoting the doctor agent learns how to treat illness[13]. These applications have demonstrated the capability of LLM agents to simulate human behaviors and interactions, providing insights for modeling more realistic urban mobility. By integrating the strengths of data-driven, contextual, and motivational models, these LLM-agent approaches aim to create high-fidelity simulations that reflect the true diversity and complexity of human mobility[22, 26]. However, the privacy protection limited achievements of the large individual mobility data with fine-grained personal profile (including age, income-level, education level, marital and parental status), thus, it is hardly to simulate human mobility behaviors in the real world.

## 3. PROBLEM DEFINTION

**Travel Diary**. An individual travel diary is donated as a set of location-based trajectory point sequences, $x = \{s_1, s_2, ..., s_n\}$, where each point $s_i$ can be expressed as ($t_i$, $l_i$, $e_i$, $d_i$, $b_i$). $l_i$ denotes the spatial location, either in the form of region ID or coordinate of latitude and longitude, $t_i$ refers to the visit time. $e_i$, $d_i$, and $b_i$ represent the travel purpose, travel distance, and travel mode, respectively.

**Individual Profiles**. There are 10 categories individual attributes in this study, such as *age*, *gender*, *occupation*, *income*, *education*, *car statue*, *housing statue*, *home address*, *primary transportation*, and *work address*, presented as $u_i = \{a_1, a_2, ... a_{10}\}$. And individuals are divided into fine-grained group according to the multiscale combination of their attributes.

**Travel Diary Generation.** Given a real-world travel survey dataset, ($X = \{x_1, x_2, ... x_N\}$ and $U = \{u_1, u_2, ... u_i\}$). The goal of the mission is to learn realistic individual mobility patterns to generate new travel diary data $Y = \{y_1, y_2, ...y_i\}$ that has similar characteristics and mobility patterns as $X$.

## 4. METHOD

Individual's mobility decisions can be inferred according to the historical mobility pattern and motivations under specific situations or events. Our framework composes two progressive stages shown in Figure 2: *Understanding-based Mobility Pattern Extraction* and *Reasoning-Based Travel Diary Generation*.

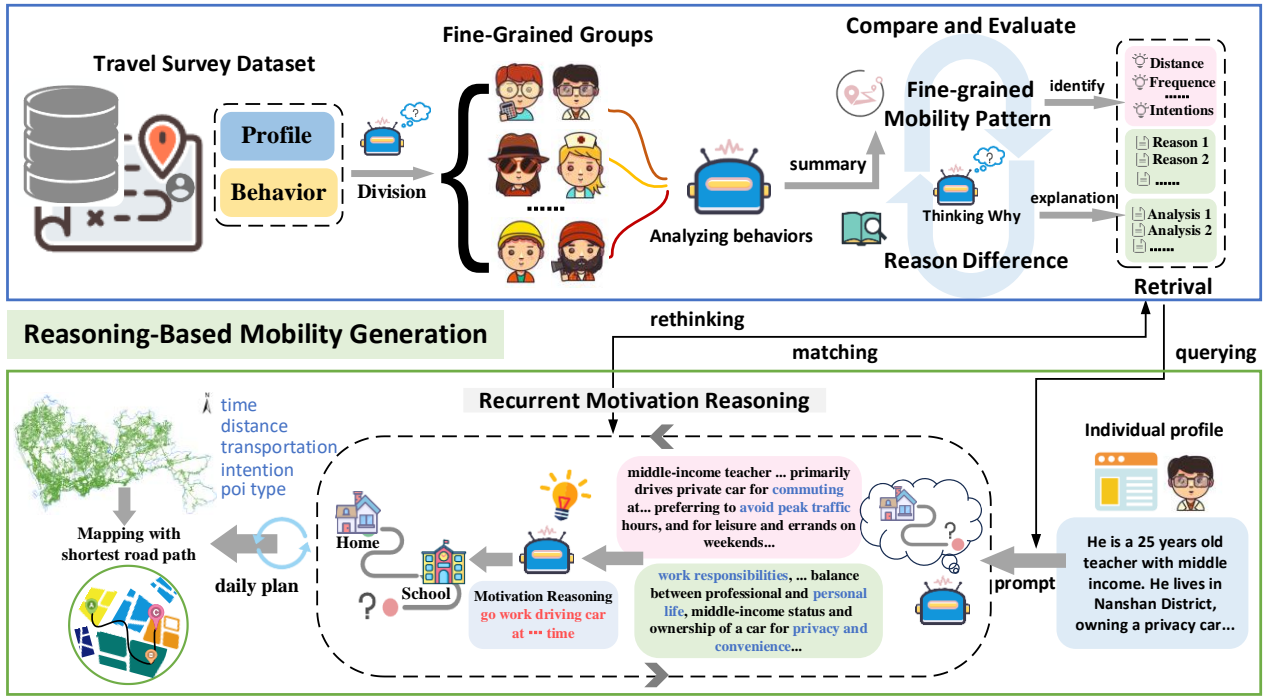

Figure 2. Workflow of the MobAgent.

## 4.1 Understanding-Based Mobility Pattern Extraction

Explaining why mobility happens, and learning the behavioral difference in different groups allows LLM agents accurately understand and infer underlying mobility motivation. We extract fine-grained mobility patterns and generate explanation to construct contextual knowledge of human mobility in the real world.

### 4.1.1 Group division and Pattern Identification

To capture the complex diversity in human mobility behaviors, we first divide the individuals into fine-grained groups based on multiple individual profiles, and then subgroups are created through hierarchical clustering. For instance, within the 'teacher' attribute, we differentiate between various age range, income levels, and education backgrounds. This allows us to observe more specific and personable mobility behaviors and avoid the pitfalls of averaging out individual differences. For each subgroup, LLM is employed to preliminarily extract these patterns, focusing on key aspects such as travel time, frequency, preferred travel modes, intentions, and common travel destinations.

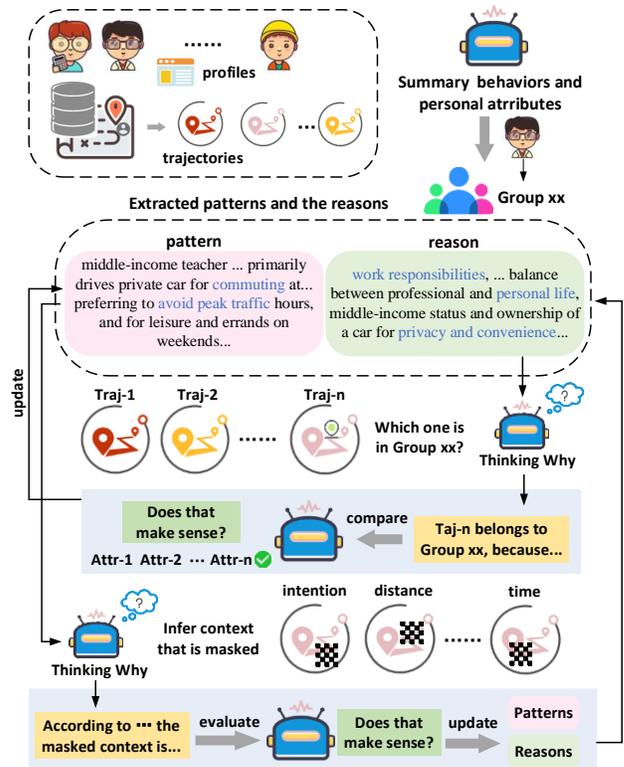

Figure 3. understanding-based patterns extraction

### 4.1.2 Pattern Comparison and Self-Evaluation

We conduct the contextual analysis and pattern comparison to understand the underlying reasons behind observed mobility patterns i.e., general patterns for human mobility, and personal variations. At macro level, the general patterns include the commute needs, constrained travel distance and frequency in specific spatiotemporal frames. At micro level, personal mobility comparison aims to explain how individual attributes, like gender and income, influence mobility behaviors such as travel purposes and timing. For example, higher-income individuals might prefer private transportation and travel longer distances for leisure activities; gender differences can reveal that women, particularly in certain age groups, may have distinct travel patterns due to childcare responsibilities or safety concerns at specific times of the day.

As shown in Figure 3, we construct a self-evaluation workflow based on the chain-of-thought (CoT) mechanism. Specifically, given the extracted mobility patterns as prior context, we select anonymized personal trajectories to let the LLM infers the group they belonged to; and the LLM completes the masked information in personal travel diaries, such as the timing or intentions. In this process, the LLM recursively evaluate and update the hierarchies and mobility patterns of individuals, by simulating mobility decision process and explain the difference of mobility behaviors from the aspect of hierarchical attributes.

### 4.2 Reasoning-Based Travel Diary Generation

Retrieval augmentation provides additional information that aids LLM in more effectively responding to queries. Not only providing mobility patterns as the static background context for fine-grained groups, we also add the analysis and result of CoT achieved in section 4.1.2 to construct the cognition of mobility mechanisms. This strategy instructs the LLM to simulate a designated individual according to existed patterns with a human-like thought linking personal mobilities with extra or inner factors.

### 4.2.1 Motivation Reasoning

The LLM agent begins by examining the motivations driving individual mobility, such as commuting, attending school, shopping, leisure activities, and social interactions. Integrating individual's profile with its retrieval information i.e., mobility patterns and prior thoughts as a complete prompt, the agent can initially construct an individual's daily plan. In the motivation reasoning process, the agent considers the intention, time, location category, distance, and travel mode for each travel activity. It then compares this with previously analyzed information to refine the generated motivation through a recursive learning process. For instance, when generating travel diary for a 38-year-old high-income female teacher with a randomly assigned home location. Based on the retrieval information that "*women with greater earnings or longer working years, who are typically not bound by the constraints of clocking in and out of work…*", the agent infers from her occupation and age that she prefers to avoid rush hour traffic on weekdays and visits family-friendly locations on weekends. Her high-income status influences her choice of travel mode and shopping destinations.

Through the recursive thinking and reasoning process, the agent continually refines its output by comparing generated trajectories against real-world situations.

### 4.2.2 Mapping to Physical Location with the Shortest Path of Road Network

There are two commonly used approaches to match the generated semantic mobility with the geo-location in a real urban environment: directly predict geographic coordinates in the generation process[26] and correspond travel intentions with real POI location by mechanistic model[22]. However, human movement is physically constrained by and influenced by the physical spatial structure, particularly the road network. Benefited from considering the attributes such as socioeconomic status, personal preferences, and lifestyles in the pattern extraction part, our framework generates more realistic individual mobility motivations.

The inferred travel distance $d_i$, can support us to find the geo-location accurately matched with the intention $e_i$ through the shortest path between blocks based on the road network; i.e., $R(d_i, e_i) \to l_{e_i}$, $R(\cdot)$ is the search function, $l_{e_i}$ is the geographic location. Thus, the generated mobility serves as a useful reflection of how the geographic layout of cities affects the mobility patterns of individuals within various fine-grained groups, so supporting real-world urban spatial design.

## 5. EXPERIMENTS

### 5.1 Experimental Setup

*Datasets.* We validate our framework on a city-wide travel survey dataset with 25,481 users and 199,380 records collected in Shenzhen, one of the largest megacities in China. The residents recorded their daily travel details in a travel surveying APP. This dataset contains 10 categories of individual attributes and 11 categories of travel details. Details of the dataset is presented in Appendix.

**Evaluation Metrics**. We evaluate generated mobility behaviors based on two dimensions: statistical evaluation and multiscale attributes evaluation.

**Statistical evaluation**. we calculate the Jensen–Shannon divergence (JSD) between the generated data and the real data in four metrics:

Step Distance (SD): measures distance between two consecutive locations in a trajectory to evaluate the spatial pattern of mobilities.

Step Interval (SI): measures time interval between two successive locations on an individual's trajectory to evaluate the temporal pattern of mobilities.

ST-LOC: calculates the number of visited Origin-Destination (OD), presented as (origin destination and time frame), to evaluate the spatial-temporal pattern of mobility.

DailyLoc: calculates the number of different locations visited per day for each of the user to evaluate the visit frequency pattern of mobility.

**Multiscale attribute evaluation.** This is conducted on groups with different attributes grains to compare the statistical evaluation. For example, the multi-grained attributes of group include (age), (age, income), (age, occupation), and (age, income, occupation), which are abbreviated as the A, A+I, A+O, and A+I+O in Table 2, respectively. The generated trajectories are aggregated on weekdays and weekends with the hourly time frame.

### 5.2 Experiments and Result

Our framework is designed in zero-shot. The pretrained LLM divided individuals into fine-grained hierarchies according to their multiple individual's attributes, and then extract mobility patterns while providing analysis under its CoT as the prior knowledge. LLM agents generate travel diaries only according to the provided individual attributes and prior knowledge.

**Table 1**. Performance comparison on statistical evaluation. MobAgent (w/o R) and MobAgent (w/o P) refer to the framework without reasoning-based strategy and understanding based patterns identification.

| Models | SD↓ | SI↓ | ST-LOC↓ | DailyLoc↓ |
| --- | --- | --- | --- | --- |
| DeepMove | 0.5614 | 0.4462 | 0.5917 | 0.5490 |
| MobAgent (w/o U) | 0.0847 | 0.1360 | 0.1304 | 0.1251 |
| MobAgent (w/o P) | 0.0811 | 0.1190 | 0.1172 | 0.1102 |
| MobAgent | 0.0658 | 0.1177 | 0.0903 | 0.0862 |

**Table 2**. Performance comparison on multiscale attribute evaluation.

| Pattern | SD↓ | SI↓ | ST-LOC↓ | DailyLoc↓ |
| --- | --- | --- | --- | --- |
| A+I+O | 0.1022 | 0.0673 | 0.0918 | 0.0898 |
| A+I | 0.1254 | 0.0819 | 0.0934 | 0.0913 |
| A+O | 0.2178 | 0.0902 | 0.1103 | 0.1455 |
| I+O | 0.1643 | 0.1561 | 0.1276 | 0.1637 |

#### 5.2.1 Over performance (RQ 1, RQ 2)

Results of evaluations on the statistical and semantic dimension are shown in Table 1 – Table 2, respectively. Comparing with the abolition models in Table 1, it can be observed that the performance of our model decreases when we remove the designed pattern extraction mechanism and the recursive reasoning strategy, which proves the effectiveness of these two components for the model to generate realistic trajectories. Furthery, it is shown that the model failing to grasp the fundamental principles of why and how people move (i.e., the individual attributes, preferences, and contextual factors driving their movements) performed worse than those that do not utilize recursive thinking for generating the next motivation in mobility sequences. This implies that understanding the "why" behind mobility is the prerequisite of refining the "what" in a step-by-step manner.

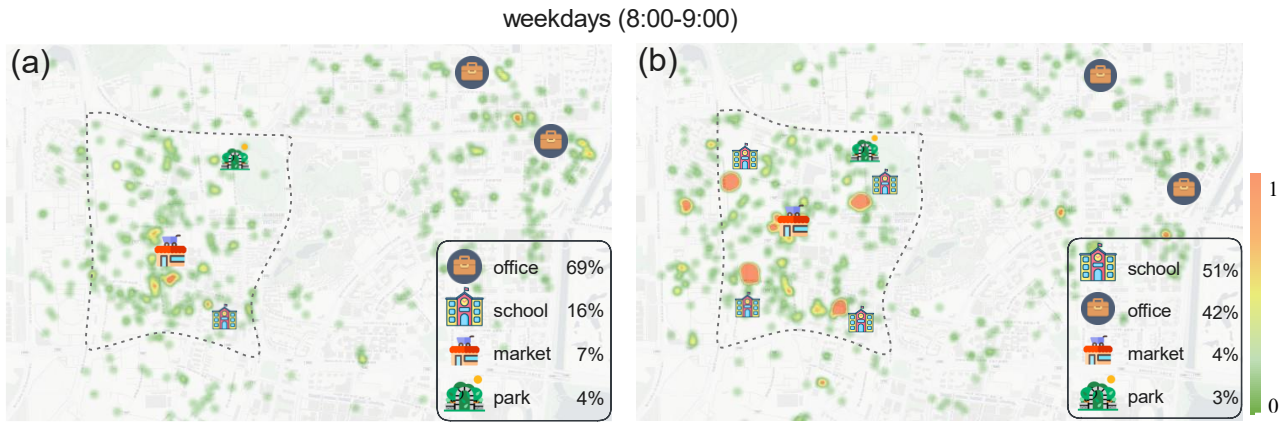

Figure 4. Comparison of activity intention of women aged at 30 to 35.

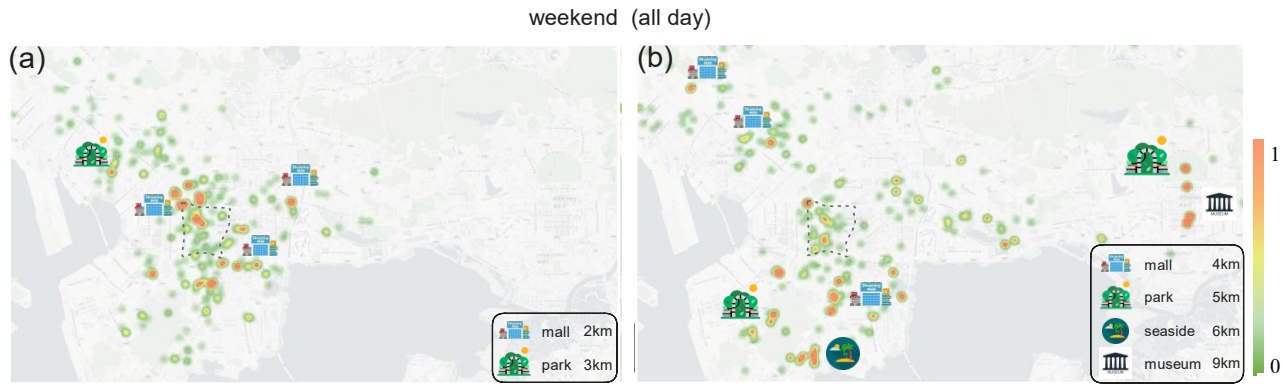

Figure 5. Comparison of the travel distance of families with the middle-high income level.

### 5.2.2 Multiscale attribute evaluation (RQ 3)

To investigate the impact of individual attributes on the effectiveness of individual mobility generation, we conducted in-depth ablation experiments on the *office staffs* group divided by different attributes. In which we focused on three core dimensions of patterns: *age*, *income*, and *occupation*. The evaluation metrics are detailed in the Table 2. It is found that when the input prompt integrates all three dimensions *age*, *income*, and *occupation*, the generated data most closely resemble real-world scenarios, demonstrating optimal performance. Following this, the *age* alone significantly influences, likely due to the distinct mobility patterns exhibited by different age groups. The *income* also plays a crucial role, particularly in determining travel distance and destination choices. In contrast, the impact of *occupation* is relatively limited in this simulation, indicating that the coarse attribute of *occupation* is limited to reflect the nuanced diversity in mobility behaviors.

In summary, *age*, *income*, and *occupation* have been proven to be highly correlated with personalized mobility patterns, with *age* having the most pronounced impact. These findings highlight the significance of finely defining and dividing personal attributes in realistic mobility behaviors generation.

### 5.2.3 Spatial-temporal Analysis (RQ 2, RQ 3)

To furtherly validate the effectiveness of the fine-grained mobility patterns extracted in Section 4.1, and explore how individuals' attributes affect their mobility choices over space and time scales. We conducted two specific experiments as illustrative examples, i.e., EXP 1 and EXP 2.

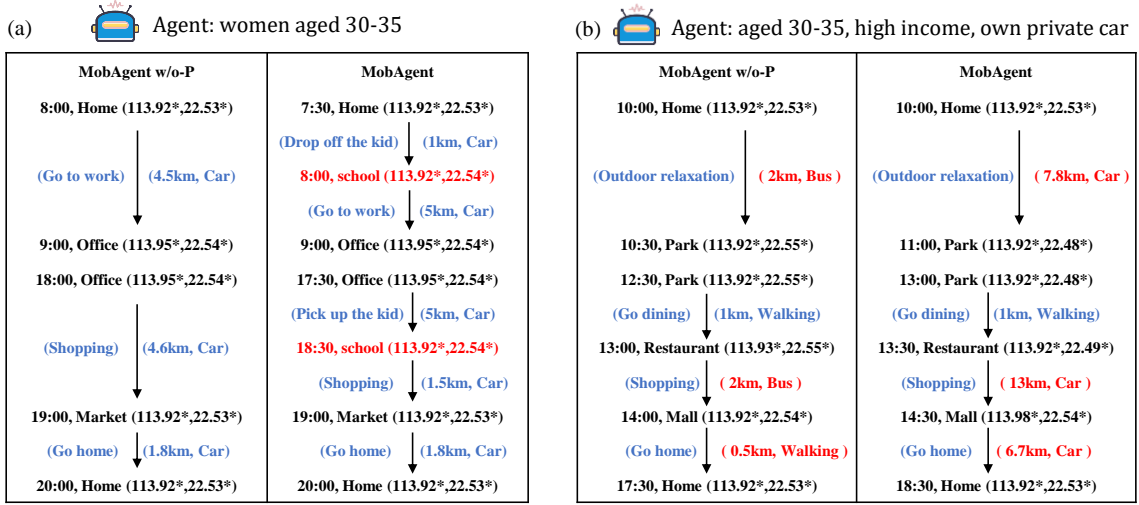

**Figure 6**. Examples of generated individual travel diary. *as privacy issues, we cover the last decimal place of location coordinate with\*.*

**EXP 1**: We designed the experiment with 200 agents, modeled as women aged 30-35 living in a specific area. Figure 4 shows the spatial distribution of their destinations during a typical workday morning from 8:00 to 9:00 AM. In Figure 4-(a), agents without the guidance of fine-grained patterns displayed a broad distribution that primarily concentrated in the business district on the left side of the map, only simply simulated the commuting motivation to work. In contrast, in Figure 4-(b), the areas around schools had significantly increased activity. That is because, agents have learned that middle-aged women often juggle both family and work responsibilities, as shown in Figure 6-(a). Their morning travels not only include commuting to work but also taking children to school and other family obligations. It validates the effectiveness of our MobAgent in capturing the relationships between the (*age, and gender*) and their travel diaries.

**EXP 2**: We generated the weekend trips of 200 male individuals, they are characterized as car owners aged 30 to 35, having middle to high-income family backgrounds. As displayed in Figure 5-(a), without the guidance of patterns, their activities tended to be centered around their residential areas, with few long-distance trips, aligning with the general preference for nearby leisure activities. In Figure 5-(b) there was a marked increase in long-distance trips, primarily to natural parks, cultural museums, seaside leisure areas, and large shopping centers on the left side of the map, which became popular choices for their leisure activities. Indicating that agents learned that the reality that high income level and the convenience of private cars often prefer planning longer trips on weekends, like the example shown in Figure 6-(b).

## 6. CONCLUSIONS

In this study, we propose a new framework for generating realistic individual travel diaries using LLM agents (MobAgent) and 0.2 million individual survey data. We emphasize understanding-based and motivation-based mechanisms to facilitate LLM to capture the fundamentals and dependencies behind mobility and individual attributes. MobAgent enables insight to finer-grained and personalized mobility patterns confirming to the reality, and generate mobility data with zero-shot learning.

Experimental results demonstrate that ignoring individual profiles would lead to the "average person" problem [11], which obscures and misleads personal mobility variations. While our framework can effectively generate more realistic individual travel diaries, especially, those who have same occupation but subtle differences in gender, age, and income level. Because this framework enables LLM to learn individual profiles at finer-grained and find associations between individual characters and their mobility behaviors. These findings demonstrate that integrating detailed individual and travel data into agent models can significantly improve the accuracy of mobility and travel behavior predictions, making simulations more reflective of real-life scenarios.

## ACKNOWLEDGEMENTS

This work was supported by the National Natural Science Foundation of China (Grant No. 42171449).

# APPENDIX

## Dataset

This dataset originates from a survey questionnaire on social networking platforms, aimed at collecting information on volunteers' travel trajectories. Participants proactively provided their travel records, and the questionnaire also included detailed personal background information, such as age, gender, occupation, and income level, to ensure the comprehensiveness and depth of the data.

**Table: Distribution of users' profiles in the dataset**

| User Profile | Category/ Coordinates |
|---|---|
| Age | <18(2.7%), 18-25(27.2%), 26-30(31.4%) 31-35(21.7%), 36-40(9.3%), 41-45(4.6%) 46-50(1.8%), >51(1.3%) |
| Gender | Male(52%), Female(48%) |
| Occupation | Management of Government Agencies(8.84%), Enterprises, and Public Institutions(8.96%), Professional and Technical Personnel(17.33%), Civil Servants and Operational Staff in Firefighting, Postal, and Telecommunications Services(4.55%), Students(9.68%), Commercial and Service Industry Personnel(21.1%), Skilled Workers(14.25%), Self-employed Individuals(7.48%), Retired/Unemployed(1.8%), Others(14.85%) |
| Income | Low (39.96%), Relatively Low (30.79%) Medium (16.87%), Relatively High (8.8%) High (3.58%) |
| Education | Bachelor's Degree (32.21%), Associate Degree (28.72%), High School Diploma (14.23%), Technical School Diploma (10.55%), Master's Degree (7.53%), Junior High School Diploma (6.01%), Primary School Diploma (0.75%) |
| Own a Car | Yes(22.32%), No(77.68%) |
| Living Situation | Rented House(47.62%), Owned House(34.69%), Dormitory(13.5%), Others(4.19%) |
| Primary Mode of Transportation | Bus and Subway(58.17%), Driving(9.95%), Taxi/Ride-Hailing(4.8%), Electric Bike/Bicycle(10.45%), Walking(15.77%), Other(0.86%) |
| Residential GPS Coordinates | Latitude, Longitude |
| Company's GPS Coordinates | Latitude, Longitude |

**Table: Basic information about the dataset**

| City | Shenzhen |
|---|---|
| Duration | November 15, 2016 - January 16, 2017 |
| Users | 25,481 |
| Records | 199,380 |
| Location Point | 32,877 |

**Table: The main fields of trajectory information records in the dataset**

| Field | Description |
|---|---|
| Travel Date | a specific date |
| Origin Address | the specific address of the starting point |
| Origin GPS Coordinates | specific latitude and longitude coordinates |
| Destination Address | the specific address of the destination |
| Destination GPS Coordinates | specific latitude and longitude coordinates |
| Travel Start Time | a specific time |
| Travel End Time | a specific time |
| Travel Duration | duration (minutes) |
| Travel Distance | distance (meters) |
| Travel Mode | Bus and Subway, Driving, Taxi/Ride-Hailing, Electric Bike/Bicycle, Walking, Other |
| Travel Purpose | Commuting to Work, Going to School, Entertainment/Dining, Medical Appointment, Picking Up/Dropping Off Someone, Returning Home, Shopping, Business Trip, Visiting Friends, Other |

# Prompt

### Initial Group Division

You have a travel survey dataset regarding travel conditions. Preliminary data analysis has revealed several key insights <INPUT 1>.
Your goal is to capture the complex diversity in human mobility behaviors present in these data across different groups, aiming to uncover the subtle differences and unique aspects of travel patterns among various subgroups. Currently, you have initially classified the subjects based on several criteria <INPUT 2>. To gain a more refined understanding of the diversity in travel patterns, please evaluate whether it is necessary to further segment the groups based on another specific dimension <INPUT 3>, considering the statistical distribution characteristics of the current data and general knowledge from daily life. Based on this consideration, please provide a rating (ranging from 1 to 10), where a higher score (closer to 10) indicates a strong recommendation for additional segmentation to obtain deeper insights.

### Fine-grained Patterns Extraction

The profiles of this group are as follows <INPUT 1>.
The patterns and distributions of their travel behaviors are as follows <INPUT 2>.
Please use common knowledge and logical reasoning to deeply consider the impact of each dimension of the group's attributes on their travel patterns. Please list in detail the correlations between each attribute and travel behavior, and analyze the fundamental reasons behind these connections.
Please first analyze the dimension <INPUT 3>.

### Patterns Update – STEP1

These are the mobility patterns you have identified <INPUT 1>. These are the travel trajectories I have provided <INPUT 2>. Please analyze the observed anonymous travel trajectories and compare them with the patterns you have identified.
Please follow the steps below:
Look at each trajectory and compare it to the patterns you identified.
Decide which group the trajectory belongs to. For example, if the trajectory shows regular long-distance travel during work hours, it might belong to 'young professionals'. Furthermore, you need to infer the age, income level, and gender of 'young professionals'.
Explain your reasoning. For instance, "This trajectory shows long commutes in the morning and evening, matching the 'young professionals' pattern."
Start with the first trajectory: Describe it, infer the group, and explain your reasoning.

### Patterns Update – STEP2

These are the mobility patterns you have identified <INPUT 1>. This is a trajectory with some information masked <INPUT 2>.
Please follow the steps below:
Fill in missing details like travel times and reasons. For example, if you know someone is a 'student' and the trajectory is missing the reason for travel at 8 AM, infer it's likely for school.
Think about why they travel the way they do. For instance, 'students' travel early for classes.
Update the patterns if needed. If you find a new behavior that doesn't fit existing patterns, refine your groups.
Start with the first anonymized trajectory: Fill in missing details, explain the reasoning, and update the patterns if necessary.

### Daily Plan Generation

Role:
You are a daily planning expert, skilled at creating clear and reasonable daily plans for individuals based on various factors (such as location, external conditions, personal attributes, and previous reflections).
Skill:
Creating a daily life plan for someone.
The individual profiles are as follows: <INPUT 1>.
The mobility patterns of this group are as follows: <INPUT 2>.
Please deeply analyze the mobility patterns and accurately infer what activities the person will likely engage in throughout the day and for how long. Present the plan in a step-by-step, executable format.
Constraints:
The plan should not include any speculative or uncertain terms.
If there are any unspecified contexts, you may make appropriate assumptions to ensure the plan appears realistic.
Provide the plan directly, without explanations or summaries.
Now, please provide the plan for <INPUT 3>.

### Recursive Reasoning of Activity Motivation

Current time: <INPUT 1>
Overall plan for today: <INPUT 2>
Earlier schedule for today: <INPUT 3>
Individual profiles: <INPUT 4>
Mobility patterns of this group: <INPUT 5>

Task: Based on the overall plan for today and the given earlier schedule, infer what the person is most likely doing at the current time. Please fully consider real-world plausibility. If there is any travel involved, provide the destination and the most likely range of travel distance, taking into account the group's attributes and mobility patterns.